\documentstyle[12pt]{article}
\newcommand{\reff}[1]{eq.~(\ref{#1})}

\newcommand{\beq}{\begin{equation}}
\newcommand{\eeq}{\end{equation}}
\newcommand{\bea}{\begin{eqnarray}} 
\newcommand{\eea}{\end{eqnarray}}
\newcommand{\bq}{\begin{quote}}
\newcommand{\eq}{\end{quote}}

\def\a{\alpha}
\def\b{\beta}
\def\d{\delta}

\def\g{\gamma}

\def\s{\sigma}

\def\beq{\begin{equation}}
\def\eeq{\end{equation}}
\def\bea{\begin{eqnarray}}
\def\eea{\end{eqnarray}}
\def\ba{\begin{array}}
\def\ea{\end{array}}
\def\no{\nonumber\\}
\def\le{\langle}
\def\re{\rangle}
\def\lt{\left}
\def\rt{\right}

\newcommand{\sect}[1]{\setcounter{equation}{0}\section{#1}}

\begin{document}
\begin{titlepage}
\begin{flushright}
UQMATH-qe2-9601\\
cond-mat/9611014
\end{flushright}
\vskip.3in
\begin{center}
{\large\bf Twisted Quantum Affine Superalgebra
$U_q[sl(2|2)^{(2)}]$, $U_q[osp(2|2)]$ Invariant R-matrices and 
a New Integrable Electronic Model}
\vskip.4in
{\large Mark D. Gould, Jon R. Links} \footnote{Australian Postdoctoral Fellow.
Email: jrl@maths.uq.oz.au}, 
{\large Yao-Zhong Zhang} \footnote{Queen Elizabeth II Fellow.
Email: yzz@maths.uq.oz.au}
\vskip.1in
{\em Department of Mathematics, University of Queensland, Brisbane,
Qld 4072, Australia}
\vskip.2in
{\large Ioannis Tsohantjis}
\vskip.1in
{\em Department of Physics, University of Tasmania, Hobart, Tas. 7001,
Australia}

\end{center}
\vskip.6in
\begin{center}
{\bf Abstract}
\end{center}
We describe the twisted affine superalgebra $sl(2|2)^{(2)}$ and its
quantized version $U_q[sl(2|2)^{(2)}]$. We investigate the tensor
product representation of the 4-dimensional grade star representation
for the fixed point subsuperalgebra $U_q[osp(2|2)]$. We work out the
tensor product decomposition explicitly and find the decomposition
is not completely reducible. Associated with this 4-dimensional grade
star representation we derive two $U_q[osp(2|2)]$ invariant R-matrices:
one of them corresponds to $U_q[sl(2|2)^{(2)}]$  and the other to
$U_q[osp(2|2)^{(1)}]$. Using the R-matrix for $U_q[sl(2|2)^{(2)}]$,
we construct a new $U_q[osp(2|2)]$ invariant strongly correlated
electronic model, which is integrable in one dimension. Interestingly,
this model reduces, in the $q=1$ limit, to the one proposed by Essler
et al which has a larger, $sl(2|2)$, symmetry.


\end{titlepage}
\newpage


\sect{Introduction\label{intro}}

Quantum affine algebras describe the underlying symmetries
of integrable systems,
conformal field theories, exactly solvable models and integrable quantum
field theories. 
Quantum affine superalgebras are ${\bf Z}_2$-graded generalizations
\cite{Kac77,Fra89} of the
bosonic quantum algebras and are mathematical objects of importance in the
study of supersymmetric theories. Examples are supersymmetric lattice
models of strongly correlated electrons 
such as the supersymmetric $t$-$J$ model \cite{Sch87},
the extended Hubbard model \cite{Ess92} and the supersymmetric $U$
model proposed in \cite{Bra95} and exactly solved in
\cite{Bed95}. In each case these models are
derived from an R-matrix satisfying the 
Yang-Baxter equation. The construction of these R-matrices can be 
achieved within the framework of the quantum affine superalgebras.

Despite their significance, quantum affine superalgebras  have so
far remained largely understudied in the literature. This is particularly
the case for the twisted quantum affine superalgebras. In this paper
we will study the twisted affine superalgebra $U_q[sl(2|2)^{(2)}]$ and
one interesting representation for its fixed point 
subsuperalgebra $U_q[osp(2|2)]$.

Lie superalgebras are much richer structures and have a more
complicated representation theory than their bosonic counterparts
\cite{Kac77,Sch77}.
For instance, a given Lie superalgebra allows many inequivalent
systems of simple roots and these give rise to different Hopf 
algebras upon deformation. As will be seen below, one has to work
with the non-standard simple root system of $sl(2|2)$ to obtain
the twisted superalgebra $sl(2|2)^{(2)}$. 

For every pair of finite-dimensional irreps
of a quantum affine superalgebra there exists a solution to the
Yang-Baxter equation \cite{Jim85}. 
In a recent paper \cite{Del96} we showed how to construct 
R-matrices for twisted bosonic quantum algebras. Our work has
immediately been taken up and generalized by the authors
in \cite{Gan96}.
For the type of irrep considered in the present paper,
however, care must be taken since the tensor product decomposition
of two such irreps is not completely reducible. This problem is
solved by introducing a nilpotent operator of order 2. Using this
approach, we will determine the spectral dependent R-matrices for
$U_q[sl(2|2)^{(2)}]$ and $U_q[osp(2|2)^{(1)}]$. 

The R-matrix for $U_q[sl(2|2)^{(2)}]$  has an interesting feature 
that in the rational limit it becomes $sl(2|2)$ invariant.  
Using this R-matrix, we will derive a new $U_q[osp(2|2)]$ invariant 
model of strongly correlated electrons which is integrable on a one 
dimension lattice. This model has different interaction terms from 
the ones in the models \cite{Sch87,Ess92,Bra95}.

This paper is organized as follows. In section \ref{twisted-sl22} and
section \ref{twisted-sl22-q}, we study the twisted 
affine superalgebra $sl(2|2)^{(2)}$ and its quantized version
$U_q[sl(2|2)^{(2)}]$, respectively. The tensor product representation of the
4-dimensional grade star representation for the fixed subsuperalgebra
$U_q[osp(2|2)]$ is also investigated in details, and basis and its dual
for this irrep are constructed explicitly.
In section \ref{r-matrix}, we derive the
R-matrix associated with the 4-dimensional irrep of
$U_q[sl(2|2)^{(2)}]$. Using this R-matrix, we propose, in section
\ref{electrons}, a new model of strongly correlated electrons which
is exactly solvable on a one-dimensional lattice. In section
\ref{untwisted-osp22-q}, we rederive the R-matrix associated with
$U_q[osp(2|2)^{(1)}]$. In section \ref{concl} we give some concluding
remarks.


\sect{Twisted Affine Superalgebra $sl(2|2)^{(2)}$\label{twisted-sl22}}

We recall the relevant information about twisted affine
superalgebras \cite{Kac77}.
Let $L$ be a finite dimensional simple Lie superalgebra and
$\tau$ a diagram automorphism 
of $L$ of order $k$. Associated to these one constructs the
twisted affine superalgebra $L^{(k)}$. In this paper we will 
assume $k=2$.
Let $L_0$ be
the fixed point subalgebra under the diagram automorphism $\s$.
We recall that
\beq\label{liesplit}
L=L_0\oplus L_1,~~~~
[L_i,L_j]=L_{(i+j) \rm{mod}2}.
\eeq
$L_1$ gives rise to a $L_0$-module under
the adjoint action of $L_0$. 
Let $\theta_0$ be its highest weight.

Let us consider $L=sl(2|2)$, whose generators we denote as $E^i_j,~i,j=1,
2,3,4$. We choose the grading $[1]=[4]=0$ and $[2]=[3]=1$. The $sl(2|2)$
generators satisfy the graded commutation relations
\beq
[E^i_j, E^k_l]=\d^k_j E^i_l-(-1)^{([i]+[j])([k]+[l])}\d^i_l E^k_j.
\eeq
We work with the non-standard root system of $sl(2|2)$. Then the associated
Dynkin diagram has an automorphism $\tau$ of order 2 \cite{Fra89}. 
Under $\tau$, the
root vectors associated to this diagram, which are $E^1_3,~E^3_2$ and 
$E^2_4$, transform in the following fashion:
\beq
\tau(E^1_3)=E^2_4,~~~~\tau(E^3_2)=E^3_2,~~~~\tau(E^2_4)=E^1_3.\label{auto1}
\eeq
This, together with relations
\bea
&&\tau(E^1_1)=-E^4_4,~~~~\tau(E^2_2)=-E^3_3,\no
&&\tau(E^3_3)=-E^2_2,~~~~\tau(E^4_4)=-E^1_1,
\eea
leads us to define the following transformation rules for other
generators in order that the
graded commutation relations are invariant under the automorphism:
\bea
&&\tau(E^1_2)=-E^3_4,~~~\tau(E^1_4)=-E^1_4,~~~\tau(E^3_4)=-E^1_2,\no
&&\tau(E^2_3)=E^2_3,~~~\tau(E^3_1)=-E^4_2,~~~\tau(E^4_2)=-E^3_1,\no
&&\tau(E^2_1)=E^4_3,~~~~\tau(E^4_1)=-E^4_1,~~~~\tau(E^4_3)=E^2_1.
   \label{auto2}
\eea
With respect to the eigenvectors of $\tau$ we have the decomposition
$sl(2|2)=sl(2|2)_0\oplus sl(2|2)_1$, where
\bea
sl(2|2)_0&=&\{X\in sl(2|2),\tau(X)=X\}=\lt\{E^2_2-E^3_3, E^2_3, E^3_2,
   \frac{1}{2}(E^3_3+E^4_4-E^1_1-E^2_2),\rt .\no
& & \lt .\frac{1}{\sqrt{2}}(-E^1_2+E^3_4),
   \frac{1}{\sqrt{2}}(E^2_1+E^4_3),\frac{1}{\sqrt{2}}(E^1_3+E^2_4),
   \frac{1}{\sqrt{2}}(-E^3_1+E^4_2)\rt\}\no
sl(2|2)_1&=&\{X\in sl(2|2),\tau(X)=-X\}=\lt\{iE^1_4, iE^4_1,
   \frac{1}{\sqrt{2}}(E^1_2+E^3_4),
   \frac{1}{\sqrt{2}}(-E^2_1+E^4_3),\rt .\no
& & \lt .\frac{1}{\sqrt{2}}(-E^1_3+E^2_4),
   \frac{1}{\sqrt{2}}(E^3_1+E^4_2), E^1_1+E^2_2+E^3_3+E^4_4\rt\}.
   \label{t-sl22}
\eea
It is easily seen that the fixed point subsuperalgebra $sl(2|2)_0$ 
is nothing but $osp(2|2)=sl(2|1)$.

We recall that $sl(2|2)$ admits Chevalley
generators $\{E_i, F_i, H_i,~i=0,1,2\}$:
\bea
&&E_1=E^2_3,~~~~F_1=E^3_2,~~~~H_1=E^2_2-E^3_3,\no
&&E_2=\frac{1}{\sqrt{2}}(-E^1_2+E^3_4),~~F_2=\frac{1}{\sqrt{2}}(E^2_1+E^4_3),
      ~~H_2=\frac{1}{2}(E^3_3+E^4_4-E^1_1-E^2_2),\no
&&E_0=i E^4_1,~~~~F_0=i E^1_4,~~~~H_0=E^1_1-E^4_4.
\eea
Here $E_i,F_i,H_i,~i=1,2,$ form the Chevalley generators
for $sl(2|2)_0$. 
$E_0\in sl(2|2)_1$ corresponds to the minimal weight vector and thus has
weight $-\theta_0$. It follows that
$H_0=-n_1 H_1-n_2 H_2$ lies in the
Cartan subalgebra $H$ of $sl(2|2)_0$. The integers $n_1,~n_2$ are known as
the Kac labels of $sl(2|2)^{(2)}$.

We now introduce the corresponding twisted affine superalgebra
$sl(2|2)^{(2)}$ which admits the decomposition
\beq
sl(2|2)^{(2)}=\bigoplus_{m\in \frac{1}{2}{\bf Z}}L_m
\oplus {\bf C} c_0,~~~~
L_m=\left\{
\begin{array}{l}
L_0(m),~m\in {\bf Z}\\
L_1(m),~m\in {\bf Z}+\frac{1}{2}
\end{array}\right.
\eeq
with $L_a(m)=\{X(m)|x\in L_a\},~a=0,1$
and $c_0$ a central charge. 
The graded Lie bracket is given by
\beq
[X(m),Y(n)]=[X,Y](m+n)+m~c_0~\delta_{m+n,0}~(X,Y),~~~
[c_0,X(m)]=0.
\eeq
Here $(~,~)$ is the fixed invariant
bilinear form on $sl(2|2)$. 
A suitable set of generators for $sl(2|2)^{(2)}$ is given by
\bea\label{simpgen}
&&e_i=E_i(0),~~~h_i=H_i(0),~~~f_i=F_i(0),~~~ i=1,2,
\no
&&e_0=E_0(1/2),~~~h_0=H_0(0)+c_0/2,~~~f_0=F_0(-1/2).
\eea
These simple generators satisfy the defining relations of
$sl(2|2)^{(2)}$:
\bea
&& [e_i,f_j]=\delta_{ij}h_i,~~~~e_2^2=0=f_2^2,\no
&&[h_0,e_0]=-2e_0,~~~~[h_0,e_1]=0,~~~~[h_0,e_2]=e_2,\no
&&[h_1,e_0]=0,~~~~[h_1,e_1]=2e_1,~~~~[h_1,e_2]=-e_2,\no
&&[h_2,e_0]=e_0,~~~~[h_2,e_1]=-e_1,~~~~[h_2,e_2]=0,\no
&&[h_0,f_0]=2f_0,~~~~[h_0,f_1]=0,~~~~[h_0,f_2]=-f_2,\no
&&[h_1,f_0]=0,~~~~[h_1,f_1]=-2f_1,~~~~[h_1,f_2]=f_2,\no
&&[h_2,f_0]=-f_0,~~~~[h_2,f_1]=f_1,~~~~[h_2,f_2]=0,\no
&&({\rm ad}e_1)^2e_2=({\rm ad}e_0)e_1=({\rm ad}e_0)^2e_2=0,\no
&&({\rm ad}f_1)^2f_2=({\rm ad}f_0)f_1=({\rm ad}f_0)^2f_2=0.
\eea

We have an algebra homomorphism, called the
{\em evaluation map}, $ev_x:U[sl(2|2)^{(2)}]\rightarrow {\bf C}[x,x^{-1}]
\otimes U[sl(2|2)]$, with $U[sl(2|2)^{(2)}],U[sl(2|2)]$ the enveloping algebras
of $sl(2|2)^{(2)}, sl(2|2)$ respectively, given by
\beq
ev_x(X(m))=x^{2m}X,~~~ ev_x(c_0)=0,
\eeq
and extended to all of $U[sl(2|2)^{(2)}]$ in the natural way.
Thus given a finite dimensional $sl(2|2)$-module $V$ 
carrying a representation $\pi$ we have a
corresponding $sl(2|2)^{(2)}$ module $V(x)={\bf C}[x,x^{-1}]\otimes V$
carrying the {\em loop representation} $\hat{\pi}$ given by
\beq
\hat{\pi}=(1\otimes\pi)ev_x.
\eeq
Below we will see such representations of $osp(2|2)$ can be quantized
to give solutions of the Yang-Baxter equation.


\sect{$U_q[sl(2|2)^{(2)}]$\label{twisted-sl22-q}}

Corresponding to the twisted affine algebra $sl(2|2)^{(2)}$
we have the twisted quantum affine algebra $U_q[sl(2|2)^{(2)}]$ with
generators $q^{\pm h_i/2},e_i,f_i,~(i=0,1,2)$ and
defining relations
\bea
&&[e_i,f_j]=\d_{ij}\frac{q^{h_i}-q^{-h_i}}{q-q^{-1}},~~~~e_2^2=0
   =f_2^2,\no
&&q^{h_0}e_0q^{-h_0}=q^{-2}e_0,~~~~q^{h_0}e_1q^{-h_0}=e_1,
   ~~~~q^{h_0}e_2q^{-h_0}=qe_2,\no
&&q^{h_1}e_0q^{-h_1}=e_0,~~~~q^{h_1}e_1q^{-h_1}=q^2e_1,
   ~~~~q^{h_1}e_2q^{-h_1}=q^{-1}e_2,\no
&&q^{h_2}e_0q^{-h_2}=qe_0,~~~~q^{h_2}e_1q^{-h_2}=q^{-1}e_1,
   ~~~~q^{h_2}e_2q^{-h_2}=e_2,\no
&&q^{h_0}f_0q^{-h_0}=q^2f_0,~~~~q^{h_0}f_1q^{-h_0}=f_1,
   ~~~~q^{h_0}f_2q^{-h_0}=q^{-1}f_2,\no
&&q^{h_1}f_0q^{-h_1}=f_0,~~~~q^{h_1}f_1q^{-h_1}=q^{-2}f_1,
   ~~~~q^{h_1}f_2q^{-h_1}=q f_2,\no
&&q^{h_2}f_0q^{-h_2}=q^{-1}f_0,~~~~q^{h_2}f_1q^{-h_2}=q f_1,
   ~~~~q^{h_2}f_2q^{-h_2}=f_2,\no
&&e_0e_1-e_1e_0=0,~~~~e_0^2e_2+e_2e_0^2-(q+q^{-1})e_0e_2e_0=0,\no
&&e_1^2e_2+e_2e_1^2-(q+q^{-1})e_1e_2e_1=0,\no
&&f_0f_1-f_1f_0=0,~~~~f_0^2f_2+f_2f_0^2-(q+q^{-1})f_0f_2f_0=0,\no
&&f_1^2f_2+f_2f_1^2-(q+q^{-1})f_1f_2f_1=0.
\eea
Throughout this
paper we will assume that $q$ is generic, i.e. not a root of unity and
$[n]_q=(q^n-q^{-n})/(q-q^{-1})$.

The algebra $U_q[sl(2|2)^{(2)}]$ is a Hopf algebra.
The coproduct is given by
\begin{eqnarray}\label{coproduct}
&&\Delta(q^{\pm h})=q^{\pm h}\otimes q^{\pm h}\nonumber\\
&&\Delta(e_i)=e_i\otimes q^{-h_i/2}+q^{h_i/2}\otimes e_i\\
&&\Delta(f_i)=f_i\otimes q^{-h_i/2}+q^{h_i/2}\otimes f_i.\nonumber
\end{eqnarray}

We omit the formulas for the antipode and the counit. The multiplication
rule for the tensor product is defined for elements $a,b,c,d\in 
U_q[sl(2|2)^{(2)}]$ by
\beq
(a\otimes b)(c\otimes d)=(-1)^{[b][c]}(ac\otimes bd).\label{gradprod}
\eeq

The (minimal) 4--dimensional irreducible representation of $U_q[sl(2|2)]$ is
undeformed. That is the representation matrices for the
fundamental generators are the same as in the classical case.
Choosing a basis $|4\re=(0,0,0,1)^t,~|3\re=(0,0,1,0)^t,~|2\re=(0,1,0,0)^t,~
|1\re=(1,0,0,0)^t$, with $|1\re,~|4\re$ even (bosonic) and
$|2\re,~|3\re$ odd (fermionic), the representation matrices are
$E^i_j=e^i_j$, where 
$(e^i_j)^k_l=\d^{ik}\d_{jl}$. Using the $U_q[sl(2|2)]$ generators
$\{E_i,~F_i,~H_i,~i=0,1,2\}$ this representation is written as
\bea
&&E_1=e^2_3,~~~~F_1=e^3_2,~~~~H_1=e^2_2-e^3_3,\no
&&E_2=\sqrt{[1/2]_q}(-e^1_2+e^3_4),~~F_2=\sqrt{[1/2]_q}
	(e^2_1+e^4_3),
      ~~H_2=\frac{1}{2}(e^3_3+e^4_4-e^1_1-e^2_2),\no
&&E_0=i e^4_1,~~~~F_0=i e^1_4,~~~~H_0=e^1_1-e^4_4.
\eea
It can be shown that there exists an evaluation representation of
$U_q[sl(2|2)^{(2)}]$ given by 
\bea
&&e_i=E_i,~~~~f_i=F_i,~~~~h_i=H_i,~~~~i=1,2,\no
&&e_0=x\, E_0,~~~~f_0=x^{-1}\,F_0,~~~~h_0=H_0.\label{ev-q}
\eea

The 4-dimensional representation of $U_q[sl(2|2)]$
is also irreducible under the $U_q[osp(2|2)]$ subsuperalgebra. We call such a
representation $U_q[osp(2|2)]$-irreducible.
Eq.(\ref{ev-q}) implies that this irreducible 4-dimensional 
$U_q[osp(2|2)]$-module,
denoted as $V$ in what follows, is affinizable to
provide also an irreducible $U_q[sl(2|2)^{(2)}]$ representation.
As in the classical case \cite{Sch77}, 
the tensor product of two such $U_q[osp(2|2)]$-irreducible
representations is not completely reducible. This can be seen as
follows. Introduce the graded permutation operator $P$ 
on the tensor product 
module $V\otimes V$ such that
\begin{equation}\label{gradperm}
P(v_\alpha\otimes v_\beta)=(-1)^{[\alpha][\beta]}
  v_\beta\otimes v_\alpha\,,~~
  \forall v_\alpha,~v_\beta\in V.
\end{equation}
We decompose the tensor product as
\beq
V\otimes V=W_+\oplus W_-
\eeq
with $W\pm$ being eigenspaces of $P$ in the $q=1$ limit:
\beq
W_\pm=\{v\in V\otimes V|\lim_{q\rightarrow 1}(P\mp 1)v=0\}.
\eeq
It is easy to check that the states
\bea
&&|\psi^-_1\re=\frac{1}{\sqrt{q^{1/2}+q^{-1/2}}}\lt(q^{1/4}
    |1\re\otimes |2\re-q^{-1/4} |2\re\otimes |1\re\rt),\no
&&|\psi^-_2\re=\frac{1}{\sqrt{q^{1/2}+q^{-1/2}}}\lt(q^{1/4}
    |1\re\otimes |3\re-q^{-1/4} |3\re\otimes |1\re\rt),\no
&&|\psi^-_3\re=|2\re\otimes |2\re,\no
&&|z\re=\frac{1}{2}(|1\re\otimes |4\re-|4\re\otimes |1\re
    +|2\re\otimes |3\re+|3\re\otimes |2\re),\no
&&|w\re=\frac{1}{\sqrt{q+q^{-1}}}\lt(q^{1/2}
    |2\re\otimes |3\re+q^{-1/2} |3\re\otimes |2\re\rt),\no
&&|\psi^-_6\re=|3\re\otimes |3\re,\no
&&|\psi^-_7\re=\frac{1}{\sqrt{q^{1/2}+q^{-1/2}}}\lt(q^{1/4}
    |2\re\otimes |4\re-q^{-1/4} |4\re\otimes |2\re\rt),\no
&&|\psi^-_8\re=\frac{1}{\sqrt{q^{1/2}+q^{-1/2}}}\lt(q^{1/4}
    |3\re\otimes |4\re-q^{-1/4} |4\re\otimes |3\re\rt)\label{state-}
\eea
span the invariant subspace $W_-$, and we set
\beq
\le\psi^-|=(|\psi^-\re)^\dagger,~~~|\psi^-\re=|\psi^-_k\re,~|z\re,~
   |w\re,
\eeq
where
\bea
&&(|\b\re\otimes |\g\re)^\dagger=(-1)^{[|\b\re][|\g\re]}
   (|\b\re)^\dagger\otimes (|\g\re)^\dagger,\no
&&(|\b\re)^\dagger=\le\b|,~~~~\forall \b=1,2,3,4.
\eea
Notice that the states $|z\re$ and $|w\re$ are not orthonormal to each
other. The remaining states are combined as follows
\bea
&&|\psi^+_1\re=|1\re\otimes |1\re,\no
&&|\psi^+_2\re=\frac{1}{\sqrt{q^{1/2}+q^{-1/2}}}\lt(q^{-1/4}
    |1\re\otimes |2\re+q^{1/4} |2\re\otimes |1\re\rt),\no
&&|\psi^+_3\re=\frac{1}{\sqrt{q^{1/2}+q^{-1/2}}}\lt(q^{-1/4}
    |1\re\otimes |3\re+q^{1/4} |3\re\otimes |1\re\rt),\no
&&|s\re=\frac{1}{\sqrt{2(q+q^{-1})}}(q^{-1/2}|1\re\otimes |4\re
    +q^{1/2}|4\re\otimes |1\re
    +q^{-1/2}|2\re\otimes |3\re-q^{1/2}|3\re\otimes |2\re),\no
&&|c\re=\frac{1}{\sqrt{2(q^2+1)(3-2q+3q^2)}}\lt ((2q^2-q+1)|1\re\otimes
    |4\re+(q^2-q+2)|4\re\otimes|1\re\rt .\no
&&~~~~~~~~~~~~~~~~~~~~~~~~~~~~~\lt .-(q+1)|2\re\otimes |3\re
    +q(q+1) |3\re\otimes |2\re\rt ),\no
&&|\psi^+_6\re=\frac{1}{\sqrt{q^{1/2}+q^{-1/2}}}\lt(q^{-1/4}
    |2\re\otimes |4\re+q^{1/4} |4\re\otimes |2\re\rt),\no
&&|\psi^+_7\re=\frac{1}{\sqrt{q^{1/2}+q^{-1/2}}}\lt(q^{-1/4}
    |3\re\otimes |4\re+q^{1/4} |4\re\otimes |3\re\rt),\no
&&|\psi^+_8\re=|4\re\otimes |4\re,
    \label{state+}
\eea
where, as above, we have used $\le\psi^+|=(|\psi^+\re)^\dagger
$, where $|\psi^+\re$ stands for $|\psi^+_k\re,~|s\re,~|c\re$.

One can show that i) the eight states (\ref{state+}) span the
invariant subspace $W_+$  and $|c\re$ is a cyclic vector for the 
corresponding representation; ii) $|s\re$ spans a 1-dimensional
invariant subspace, i.e. it is mapped into zero by all generators of
$U_q[osp(2|2)]$. However, the singlet state $|s\re$ is not separable
from the representation. Therefore the tensor product is not completely
reducible. 

Recall that $|z\re,~|w\re,~|s\re$ and $|c\re$ are not
orthonormal to each other. Let us construct the dual of these
states. Denote
\beq
|\psi_1\re\equiv |z\re,~~~|\psi_2\re\equiv |w\re,~~~
|\psi_3\re\equiv |s\re,~~~|\psi_4\re\equiv |c\re
\eeq
and define a metric $g_{ij}$:
\beq
g_{ij}=\le\psi_i|\psi_j\re,~~~~i,j=1,2,3,4.
\eeq
It is easily shown that
\bea
&&g_{11}=g_{22}=g_{33}=g_{44}=1,~~~g_{12}=g_{21}=\frac{q^{1/2}+q^{-1/2}}
     {2\sqrt{q+q^{-1}}},\no
&&g_{13}=g_{31}=\frac{q^{-1/2}-q^{1/2}}{\sqrt{2(q+q^{-1})}},~~~
    g_{23}=g_{32}=g_{24}=g_{42}=0,\no
&&g_{34}=g_{43}=0,~~~g_{14}=g_{41}=\frac{q^2-1}{\sqrt{2(q^2+1)(3-2q+3q^2)}}.
\eea
Define dual states as follows
\beq
\le\psi^i|=g^{ij}\le\psi_j|, ~~~~(g^{ij})=(g_{ij})^{-1}
\eeq
where summation on the repeated index $j$ is implied. A long exercise
leads to 
\bea
\le\psi^1|&=&\frac{2}{(1+q)^2}\lt ((1+q^2)(\le 1|\otimes \le 4|
   -\le 4|\otimes \le 1|)+(1-q)(\le 2|\otimes \le 3|-q
   \le 3|\otimes \le 2|)\rt ),\no
\le\psi^2|&=&\frac{\sqrt{1+q^2}}{1+q}\lt (-\le 1|\otimes \le 4|
   +\le 4|\otimes \le 1|-\le 2|\otimes \le 3|-
   \le 3|\otimes \le 2|\rt ),\no
\le\psi^3|&=&\frac{1}{(1+q)^2\sqrt{2(q^2+1)}}\lt ((-1+4q-q^2+2q^3)
    \le 1|\otimes \le 4|+(2-q+4q^2-q^3)\le 4|\otimes \le 1|\rt .\no
& &~~~~~~~~~~~~~\lt .-(3-2q+3q^2)(\le 2|\otimes \le 3|-q\le 3|\otimes \le 2|)
   \rt ),\no
\le\psi^4|&=&\sqrt{\frac{3-2q+3q^2}{2(q^2+1)}}\frac{1}{1+q}(\le 1|
   \otimes \le 4|+q\le 4|\otimes \le 1|+\le 2|\otimes\le 3|
   -q\le 3|\otimes \le 2|).\label{dual}
\eea

We remark that $\le\psi^4|$ spans a 1-dimensional right submodule
under the quantum group action.


\sect{R-matrix for $U_q[sl(2|2)^{(2)}]$\label{r-matrix}}

With an abuse of notation, in this section we set
$e_0=ie^4_1,~f_0=ie^1_4$ and $h_0=e^1_1-e^4_4$. It can be shown  
\cite{Jim85} that a solution to the linear equations
\begin{eqnarray}
&&R(x)\Delta(a)=\bar{\Delta}
  (a)R(x)\,,~~~\forall a\in U_q[osp(2|2)],\nonumber\\
&&R(x)\left (x\, e_0\otimes q^{-h_0/2}+
  q^{h_0/2}\otimes e_0\right )
  =\left (x\,e_0\otimes q^{h_0/2}
  +q^{-h_0/2}\otimes e_0\right )R(x)\label{r(x)1}
\end{eqnarray}
satisfies the QYBE 
\begin{equation}
R_{12}(x)R_{13}(xy)R_{23}(y)
  =R_{23}(y)R_{13}(xy)R_{12}(x).
\end{equation}
In the above, 
$\bar{\Delta}=T\cdot \Delta$, with $T$ the twist map defined by
$T(a\otimes b)=(-1)^{[a][b]}b\otimes a\,,~\forall a,b\in U_q[osp(2|2)]$ and
also, if $R(x)=\sum_ia_i\otimes
b_i$, then $R_{12}(x)=\sum_ia_i
\otimes b_i\otimes I$ etc. The solution to
(\ref{r(x)1}) is unique, up to scalar functions. The multiplicative
spectral parameter $x$ can be transformed into an additive spectral
parameter $u$ by $x=\mbox{exp}(u)$.

In all our equations we implicitly use the ``graded" multiplication rule of 
\reff{gradprod}. Thus the R-matrix of a quantum superalgebra satisfies
a ``graded" QYBE which, when written as an ordinary
matrix equation, contains extra signs:
\begin{eqnarray}
&&R(x)_{ij}^{i'j'}
  R(xy)_{i'k}^{i''k'}
  R(y)_{j'k'}^{j''k''}
  (-1)^{[i][j]+[k][i']+[k'][j']}\nonumber\\
&&~~~~~~~~~~~~~~~~=R(y)_{jk}^{j'k'}
  R(xy)_{ik'}^{i'k''}
  R(x)_{i'j'}^{i''j''}
  (-1)^{[j][k]+[k'][i]+[j'][i']}.
\end{eqnarray}
However after a redefinition
\beq\label{redef}
\tilde{R}(\cdot)_{ij}^{i'j'}
=R(\cdot)_{ij}^{i'j'}\,
(-1)^{[i][j]}
\eeq
the signs disappear from the equation. Thus any solution of the ``graded"
QYBE arising from the R-matrix of a quantum superalgebra
provides also a solution of the standard QYBE after
the redefinition in \reff{redef}.

Set
\begin{equation}
\check{R}(x)=PR(x)
\end{equation}
where $P$ is the graded permutation operator on $V\otimes V$.
Then (\ref{r(x)1}) can be rewritten as
\begin{eqnarray}
&&\check{R}(x)\Delta(a)=\Delta(a)
  \check{R}(x)\,,~~~\forall a\in U_q[osp(2|2)],\nonumber\\
&&\check{R}(x)\left (x\,e_0\otimes
  q^{-h_0/2}+q^{h_0/2}\otimes e_0\right )
  =\left (e_0\otimes q^{-h_0/2}+
  x\,q^{h_0/2}\otimes e_0\right )\check{R}(x)\label{r(x)2}
\end{eqnarray}
and in terms of $\check{R}(x)$ the QYBE becomes
\bea
&&(I\otimes\check{R}(x))(\check{R}(xy)
  \otimes I)(I\otimes\check{R}(y))
  =(\check{R}(y)\otimes I)(I\otimes \check{R}(xy))
  (\check{R}(x)\otimes I).\label{ybe}
\eea
Note that this equation,
if written in matrix form, does not have extra signs.
This is because the definition of the graded permutation operator
in \reff{gradperm} includes the signs of \reff{redef}.
In the following we will normalize the $R$-matrix
$\check{R}(x)$ in such a way that 
$\check{R}(x)\check{R}(x^{-1})=I$,
which is usually called the unitarity condition in the literature.

Let us proceed to solve $\check{R}(x)$ satisfying (\ref{r(x)2}) for
$U_q[sl(2|2)^{(2)}]$, that is for $e_0=ie^4_1,~h_0=e^1_1-e^4_4$.
As we have shown in the last section, the tensor product decomposition is not
completely reducible. Therefore the tensor product graph method
developled in \cite{Del96,Gan96} is not applicable to the present case.
Let $P[\pm]$ denote the (central) projection operators defined by
\beq
P[\pm](V\otimes V)=W_\pm
\eeq
and $N$ the operator mapping the cyclic vector of $V\otimes V$ to the singlet
$V_0\subset W_+\subset V\otimes V$. Obviously $N$ is nilpotent of
order 2 (i.e. $N^2=0$). Using the states we have found in last
section, $P[\pm]$ and $N$ can be expressed as
\bea
P[+]&=&|\psi^+_1\re\le\psi^+_1|+|\psi^+_2\re\le\psi^+_2|+
       |\psi^+_3\re\le\psi^+_3|+|\psi^+_6\re\le\psi^+_6|+\no
& &    |\psi^+_7\re\le\psi^+_7|+|\psi^+_8\re\le\psi^+_8|+
       |\psi_3\re\le\psi^3|+|\psi_4\re\le\psi^4|,\no
P[-]&=&|\psi^-_1\re\le\psi^-_1|+|\psi^-_2\re\le\psi^-_2|+
       |\psi^-_3\re\le\psi^-_3|+|\psi^-_6\re\le\psi^-_6|+\no
& &    |\psi^-_7\re\le\psi^-_7|+|\psi^-_8\re\le\psi^-_8|+
       |\psi_1\re\le\psi^1|+|\psi_2\re\le\psi^2|,\no
N&=& f(q)\;|\psi_3\re\le\psi^4|,\label{projectors}
\eea
where $f(q)$ is an arbitrary factor depending on $q$. It is worth
pointing out that $P[\pm]$ and $N$ are all quantum group $U_q[osp(2|2)]$
invariants. Moreover they satisfy the following relations
\bea
&&P[\pm]P[\pm]=P[\pm],~~~~N^2=0,~~~~P[+]P[-]=P[-]P[+]=0,\no
&&P[+]N=NP[+]=N,~~~P[-]N=NP[-]=0,~~~P[+]+P[-]=1.\label{project-r}
\eea
With the help of these operators,
the most general $\check{R}(x)$ satisfying the first
equation in (\ref{r(x)2}) may be written in the form
\begin{equation}
\check{R}(x)=\rho_+(x)\;P[+]+\rho_N(x)\;N+\rho_-(x)P[-]\label{ansatz}
\end{equation}
where $\rho_\pm(x),~\rho_N(x)$, are  unknown functions depending on
$x,~q$.  

Multiplying the second equation in (\ref{r(x)2}) by $P[+]$ from the left 
and the resulting equation by $P[+]$ from the right, one gets
\bea
&&\lt(\rho_+(x)P[+]+\rho_N(x)N\rt) \left (x\,e_0\otimes
  q^{-h_0/2}+q^{h_0/2}\otimes e_0\right )P[+]\no
&&~~~~~~~~~~~~~~~  =P[+]\left (e_0\otimes q^{-h_0/2}+
  x\,q^{h_0/2}\otimes e_0\right )\lt(\rho_+(x)P[+]+\rho_N(x)N\rt)
\eea
where (\ref{project-r}) has been used.
With the help of (\ref{projectors}), (\ref{state-}), (\ref{state+})
and (\ref{dual}), one obtaines from the above equation
\beq
\rho_N(x)=\frac{1-q}{f(q)}\frac{2(q+q^{-1})}{\sqrt{3-2q+3q^2}}
     \frac{1-x}{1+x}\rho_+(x).
\eeq

If one multiplies the second equation in (\ref{r(x)2}) by $P[-]$ from the left
and the resulting equation by $P[+]$ from the right, one has
\bea
&&\rho_-(x)P[-] \left (x\,e_0\otimes
  q^{-h_0/2}+q^{h_0/2}\otimes e_0\right )P[+]\no
&&~~~~~~~~~~~~~~~~  =P[-]\left (e_0\otimes q^{-h_0/2}+
  x\,q^{h_0/2}\otimes e_0\right )\lt(\rho_+(x)P[+]+\rho_N(x)N\rt)
\eea
which gives rise to
\beq
\rho_-(x)=\frac{1-xq}{x-q}\rho_+(x)
\eeq
It follows that 
\beq
\check{R}(x)=P[+]+
     \frac{1-q}{f(q)}\frac{2(q+q^{-1})}{\sqrt{3-2q+3q^2}}
     \frac{1-x}{1+x} N+\frac{1-xq}{x-q}P[-].\label{twisted-r}
\eeq
Remember that the arbitrary factor $f(q)$ in (\ref{twisted-r}) 
cancells out with the same
factor $f(q)$ appearing in the definition of $N$.

An interesting feature of this $U_q[osp(2|2)]$ invariant R-matrix is that
in the rational limit
the $N$ term disappears from $\check{R}(x)$ and the resultant rational
R-matrix becomes $sl(2|2)$ invariant: the 36-vertex model
reduces to a 28-vertex one in the rational limit!
We also point out that a similar $U_q[osp(2|2)]$-invariant R-matrix
has essentially been obtained in \cite{Deg90} using a different approach.
We have shown here that this R-matrix actually comes from the twisted
quantum affine superalgebra $U_q[sl(2|2)^{(2)}]$.


\sect{New $U_q[osp(2|2)]$ Invariant Electronic Model\label{electrons}}

In this section we propose a new $U_q[osp(2|2)]$-invariant strongly
correlated electronic model on the unrestricted $4^L$-dimensional
electronic Hilbert space $\otimes^L_{n=1}{\bf C}^4$, where $L$ is
the lattice length. This model has different interaction terms from 
previous ones introduced in \cite{Sch87,Ess92,Bra95}.

We recall that electrons
on a lattice are described by canonical Fermi operators $c_{i,\s}$
and $c_{i,\s}^\dagger$ satisfying the anti-commutation relations given by
$\{c_{i,\s}^\dagger, c_{j,\tau}\}=\d_{ij}\d_{\s\tau}$, where 
$i,j=1,2,\cdots,L$ and $\s,\tau=\uparrow,\;\downarrow$.
The operator $c_{i,\s}$ annihilates an electron of spin $\s$ 
at site $i$, which implies that the Fock vacuum $|0\re$ satisfies 
$c_{i,\s}|0>=0$.
At a given lattice site $i$ there are four possible electronic states:
\beq
|0\re\,,~~~|\uparrow\re_i=c_{i,\uparrow}^\dagger|0\re\,,~~~
  |\downarrow\re_i=c_{i,\downarrow}^\dagger|0\re\,,~~~
  |\uparrow\downarrow\re_i=c_{i,\downarrow}^\dagger 
  c_{i,\uparrow}^\dagger|0\re\,.\label{states}
\eeq
By $n_{i,\s}=c_{i,\s}^\dagger c_{i,\s}$ we denote the number operator for
electrons with spin $\s$ on site $i$, and we write $n_i=n_{i,\uparrow}+
n_{i,\downarrow}$. The spin operators $S\,,~S^\dagger\,,~
S^z$, (in the following, the global operator ${\cal O}$ will be always
expressed in terms of the local one ${\cal O}_i$ as 
${\cal O}=\sum_{i=1}^L{\cal O}_i$ in one dimension)
\beq
S_i=c_{i,\uparrow}^\dagger c_{i,\downarrow}\,,~~~
  S_i^\dagger=c_{i,\downarrow}^\dagger c_{i,\uparrow}\,,~~~
  S_i^z=\frac{1}{2}(n_{i,\downarrow}-n_{i,\uparrow})\,,\label{s-operator}
\eeq
form an $sl(2)$ algebra and they commute with the hamiltonians that we
consider below.

Using the R-matrix (\ref{twisted-r}) and denoting
\beq
\check{R}_{i,i+1}(x)=I\otimes\cdots I\otimes
  \underbrace{\check{R}(x)}_{i~i+1}\otimes I
  \otimes\cdots\otimes I
\eeq
one may define the local hamiltonian
\beq
H_{i,i+1}=\left .\frac{d}{dx}\check{R}_{i,i+1}
   (x)\right |_{x=1}
\eeq
By (\ref{projectors}), (\ref{state-}), (\ref{state+}) and (\ref{dual})
and choosing
\beq
|4\re\equiv |0\re,~~~~
|3\re\equiv |\uparrow\re,~~~~
|2\re\equiv |\downarrow\re,~~~~
|1\re\equiv |\uparrow\downarrow\re,\label{choice}
\eeq
one gets, after tedious but straightforward manipulation,
\bea
H&\equiv& \sum_{\le i,j\re}H_{i,j},\no
H_{i,j}&=&c^\dagger_{i,\uparrow}c_{j,\uparrow}\lt [1-n_{i,\downarrow}
   -n_{j,\downarrow}-\frac{1}{2}(q^{1/2}-q^{-1/2})\lt (n_{i,\downarrow}
   (1-n_{j,\downarrow})+n_{j,\downarrow}(1-n_{i,\downarrow})\rt)\rt]\no
& &+c^\dagger_{i,\downarrow}c_{j,\downarrow}\lt [1-n_{i,\uparrow}
   -n_{j,\uparrow}-\frac{1}{2}(q^{1/2}-q^{-1/2})\lt (q (1-n_{i,\uparrow})
   n_{j,\uparrow}+q^{-1}n_{i,\uparrow}(1-n_{j,\uparrow})\rt)\rt]\no
& &+c^\dagger_{j,\uparrow}c_{i,\uparrow}\lt [1-n_{i,\downarrow}
   -n_{j,\downarrow}+\frac{1}{2}(q^{1/2}-q^{-1/2})\lt (n_{i,\downarrow}
   (1-n_{j,\downarrow})+n_{j,\downarrow}(1-n_{i,\downarrow})\rt)\rt]\no
& &+c^\dagger_{j,\downarrow}c_{i,\downarrow}\lt [1-n_{i,\uparrow}
   -n_{j,\uparrow}+\frac{1}{2}(q^{1/2}-q^{-1/2})\lt (q (1-n_{i,\uparrow})
   n_{j,\uparrow}+q^{-1}n_{i,\uparrow}(1-n_{j,\uparrow})\rt)\rt]\no
& &+\frac{1}{2}(q^{1/2}+q^{-1/2})\lt(S^\dagger_iS_j+S^\dagger_jS_i
   -q^{-1}n_{i,\uparrow}n_{j,\downarrow}-
   q\,n_{i,\downarrow}n_{j,\uparrow}\rt)\no
& &-\frac{1}{2}(q^{1/2}+q^{-1/2})\lt (c^\dagger_{i,\uparrow}
   c^\dagger_{i,\downarrow}c_{j,\downarrow}c_{j,\uparrow}+{\rm h.c.}+
   (q-q^{-1})n_{i,\uparrow}n_{j,\uparrow}(n_{j,\downarrow}-
   n_{i,\downarrow})\rt )\no
& &+\frac{q^{-2}-2q-3}{2(q^{1/2}+q^{-1/2})}n_{i,\uparrow}n_{i,\downarrow}
   +\frac{q^2-2q^{-1}-3}{2(q^{1/2}+q^{-1/2})}n_{j,\uparrow}
   n_{j,\downarrow}\no
& &+q^{1/2}(n_{i,\uparrow}+n_{i,\downarrow})
   +q^{-1/2}(n_{j,\uparrow}+n_{j,\downarrow}),\label{hamiltonian}
\eea
where $<i,j>$ denote nearest neighour links on the lattice.
In deriving (\ref{hamiltonian}), use has been made of the following identities 
\bea
&&|0\re\le 0|+|\downarrow\re\le\downarrow|+|\uparrow\re\le\uparrow|
  +|\uparrow\downarrow\re\le\uparrow\downarrow|=1,\no
&&|\uparrow\downarrow\re\le\uparrow\downarrow|=n_\uparrow n_\downarrow,\no
&&|\uparrow\re\le\uparrow|=n_\uparrow-n_\uparrow n_\downarrow,~~~~
   |\downarrow\re\le\downarrow|=n_\downarrow-n_\uparrow n_\downarrow.
\eea

Our hamiltonian is supersymmetric and the supersymmetry algebra is
$U_q[osp(2|2)]$. The global hamiltonian commutes
with global number operators of spin up and spin down, respectively.
Moreover the model is exactly solvable on the one dimensional lattice.

In the $q=1$ limit, our model reduces to one proposed by Essler et al
\cite{Ess92} which has a larger, $sl(2|2)$, symmetry.


\sect{$U_q[osp(2|2)^{(1)}]$ R-matrix Revisited\label{untwisted-osp22-q}}

The 4-dimensional grade star irrep of $U_q[osp(2|2)]$ can
also be extended to carry an irreducible representation of 
the untwisted quantum affine superalgebra $U_q[osp(2|2)^{(1)}]$.
In this case $e_0$ and $f_0$ are odd and given by
\bea
&&e_0=\sqrt{[1/2]_q}(-e^3_1+e^4_2), ~~f_0=-\sqrt{[1/2]_q}(e^1_3+e^2_4),\no
&&h_0=-\frac{1}{2}(e^2_2+e^4_4-e^1_1-e^3_3).\label{untwisted-efh's}
\eea
Denote the  R-matrix in the
present case as $\check{R}_{\rm ut}(x)$. In principal, this R-matrix can
be obtained by carefully taking the $\a=-\frac{1}{2}$ limit of the 
corresponding R-matrix found in \cite{Bra94}. Here we rederive it
more rigorously. 

With the explicit expression (\ref{untwisted-efh's}) 
of $e_0,~f_0$ and $h_0$, and
writing the most general $\check{R}_{\rm ut}(x)$ as the form 
\begin{equation}
\check{R}_{\rm ut}(x)=\varrho_+(x)\;P[+]+\varrho_N(x)\;N+\varrho_-(x)P[-]
\end{equation}
the Jimbo equations
\begin{eqnarray}
&&\check{R}_{\rm ut}(x)\Delta(a)=\Delta(a)
  \check{R}_{\rm ut}(x)\,,~~~\forall a\in U_q[osp(2|2)],\nonumber\\
&&\check{R}_{\rm ut}(x)\left (x\,e_0\otimes
  q^{-h_0/2}+q^{h_0/2}\otimes e_0\right )
  =\left (e_0\otimes q^{-h_0/2}+
  x\,q^{h_0/2}\otimes e_0\right )\check{R}_{\rm ut}(x)\label{r(x)3}
\end{eqnarray}
can be solved by direct computations, as we did in last section.
Here we proceed a bit differently. We recall that 
the braid generator $\s$, which satisfies the first equation in
(\ref{r(x)3}) and the relation
\beq
\s\lt(e_0\otimes q^{-h_0/2}\rt)=\lt(q^{h_0/2}\otimes e_0\rt)\s,\label{braid-r}
\eeq
is given by taking the $x\rightarrow \infty$ limit of $\check{R}_{\rm
ut} (x)$
\beq
\s=\check{R}_{\rm ut}(\infty)=\varrho_+(\infty)\;P[+]+\varrho_N(\infty)
    \;N+\varrho_-(\infty)\;P[-].\label{braid}
\eeq
On the other hand, the braid generator can also be obtained by taking the
$x\rightarrow\infty$ limit of $\check{R}(x)$ in the twisted case:
\beq
\s=\check{R}(\infty)=P[+]-
     \frac{1-q}{f(q)}\frac{2(q+q^{-1})}{\sqrt{3-2q+3q^2}}N-q P[-].
\eeq
Comparing the above two $\s$'s, one gets
\beq
\varrho_+(\infty)=1,~~~~\varrho_-(\infty)=-q,~~~~\varrho_N(\infty)=-
     \frac{1-q}{f(q)}\frac{2(q+q^{-1})}{\sqrt{3-2q+3q^2}}.\label{varrho-infty}
\eeq
Multiply the second equation in (\ref{r(x)3}) by $P[+]$ from the left
and the reuslting equation by $P[-]$ from the right, one obtains
\bea
&&\lt (\varrho_+(x)P[+]+\varrho_N(x)N\rt)\left (x\,e_0\otimes
  q^{-h_0/2}+q^{h_0/2}\otimes e_0\right )P[-]\no
&&~~~~~~~~~~~~~~~~  =\varrho_-(x)P[+]\left (e_0\otimes q^{-h_0/2}+
  x\,q^{h_0/2}\otimes e_0\right )P[-]
\eea
This equation is simplified upon using  the relations
\bea
&&P[+]\lt(q^{h_0/2}\otimes e_0\rt)P[-]=\lt(\frac{\varrho_+(\infty)}
  {\varrho_-(\infty)}P[+]+\frac{\varrho_N(\infty)}{\varrho_-(\infty)}\rt)
  \lt(e_0\otimes q^{-h_0/2}\rt)P[-]\no
&&N\lt(q^{h_0/2}\otimes e_0\rt)P[-]=\frac{\varrho_+(\infty)}
  {\varrho_-(\infty)}N
  \lt(e_0\otimes q^{-h_0/2}\rt)P[-],
\eea
which are derived from (\ref{braid-r}) by multiplying 
$P[+]$ and $N$ from the left, respectively and $P[-]$ from the right.
The simplified expressions read
\bea
\lt\{\lt(1+x\frac{\varrho_+(\infty)}{\varrho_-(\infty)}\rt)\varrho_-(x)
  -\lt(x+\frac{\varrho_+(\infty)}{\varrho_-(\infty)}\rt)\varrho_+(x)\rt\}
  P[+]\lt(e_0\otimes q^{-h_0/2}\rt)P[-]&=&0,\no
\lt\{\frac{\varrho_N(\infty)}{\varrho_-(\infty)}\lt (x\varrho_-(x)
  -\varrho_+(x)\rt)-\varrho_N(x)
  \lt(x+\frac{\varrho_+(\infty)}{\varrho_-(\infty)}\rt)\rt\}
  N\lt(e_0\otimes q^{-h_0/2}\rt)P[-]&=&0.\no
\eea
One can easily show that
\bea
P[+]\lt(e_0\otimes q^{-h_0/2}\rt)P[-]&\neq&0,\no
N\lt(e_0\otimes q^{-h_0/2}\rt)P[-]&\neq&0.
\eea
It follows that
\bea
\varrho_-(x)&=&\frac{x\varrho_-(\infty)+\varrho_+(\infty}
    {\varrho_-(\infty)+x\varrho_+(\infty)}\varrho_+(x)=
    \frac{1-xq}{x-q}\rho_+(x),\no
\varrho_N(x)&=&\frac{\varrho_N(\infty)}
     {x\varrho_-(\infty)+\varrho_+(\infty)}
     \lt (x\varrho_-(x)-\varrho_+(x)\rt)\no
&=&  \frac{1-q}{f(q)}\frac{2(q^2+1)}{\sqrt{3-2q+3q^2}}
     \frac{(x-1)(x+1)}{(1-xq)(x-q)}\rho_+(x),
\eea
where (\ref{varrho-infty}) has been used. Thus
\beq
\check{R}_{\rm ut}(x)=P[+]+\frac{1-q}{f(q)}
     \frac{2(q^2+1)}{\sqrt{3-2q+3q^2}}
     \frac{(x-1)(x+1)}{(1-xq)(x-q)}N+
     \frac{1-xq}{x-q}P[-].\label{untwisted-r}
\eeq

The R-matrix (\ref{untwisted-r}) also leads to an integrable model
of strongly correlated electrons, which, up to a similarity
transformation, is the $\a=-\frac{1}{2}$ limit of the model proposed
in the second paper of the reference \cite{Bra95}.


\sect{Concluding Remarks\label{concl}}

We have described the twisted quantum affine superalgebra
$U_q[sl(2|2)^{(2)}]$ and obtained the R-matrix $\check{R}(x)$,
corresponding to the four dimensional irrep, which is invariant
under $U_q[osp(2|2)]$ where $osp(2|2)$ is the fixed point
subsuperalgebra under the automorphism on $sl(2|2)$. This leads to a
new 4-state model of strongly correlated electrons for which the local
Hamiltonian was determined explicitly. It has $U_q[osp(2|2)]$ 
invariance and the model is exactly solvable in one dimension via the
QISM. It is interesting that in the classical ($q\rightarrow 1$)
limit, the R-matrix admits $sl(2|2)$ invariance and the corresponding
exactly solvable model reduces to that Essler et al \cite{Ess92}.

It was moreover shown that the underlying 4-dimensional irrep also
gives rise to another $U_q[osp(2|2)]$-invariant R-matrix associated
with the untwisted quantum affine superalgebra $U_q[osp(2|2)^{(1)}]$.
This R-matrix was determined explicitly and also determines a
4-state model of strongly correlated electrons, exactly solvable
in one dimension. This latter model in fact arises as the
$\a=-\frac{1}{2}$ limit of the model proposed in \cite{Bra95}.

The R-matrices determined in this paper exhibit the novel feature
of having a $U_q[osp(2|2)]$-invariant nilpotent component. They
give rise to a local Hamiltonian for a quantum spin chain which is not
hermitian, but nevertheless admits real eigenvalues (for parameters
in the appropriate range). This arises due to the fact that in the
reduction of the tensor product of the 4-dimensional irrep with
itself into $U_q[osp(2|2)]$ modules, an indecomposable occurs.
New techniques are thus required for the solution of the corresponding
Jimbo equations, as we have shown in the paper. Our approach yields
a new extension of the twisted tensor product graph method introduced in
\cite{Del96}.

The R-matrices, and corresponding exactly solvable models, investigated
above are in fact the simplest in an infinite hierarchy
arising from the twisted quantum affine superalgebra
$U_q[sl(m|n=2k)^{(2)}]$. Such R-matrices all admit $U_q[osp(m|n)]$
invariance and give rise to new supersymmetric lattice models,
exactly solvable in one dimension. Their study is thus of great
interest, and it is expected that the novel features observed in
the case studies of this paper, will also occur in general. It is
hoped that the techiniques we have introduced above will provide a
basis for the explicit determination of these more general R-matrices
and their corresponding lattice models.

\vskip.3in
\begin{center}
{\bf Acknowledgements} 
\end{center}
This work has been financially supported by the Australian Research
Council.

\newpage

\end{document}